\documentclass[doublecol]{epl2} 

\usepackage{amsmath}
\usepackage{stfloats} 
\usepackage{multirow}
\usepackage{booktabs}

\title{Production of Higgs boson in association with a pair of fermions in the presence of a circularly polarized laser field}
\shorttitle{Production of a Higgs boson in association with a pair of fermions in the presence of a circularly polarized laser field}

\author{M. Ouali\inst{1} \and M. Ouhammou\inst{1} \and R. Benbrik \inst{2} \and S. Taj\inst{1} \and B. Manaut\inst{1}}
\shortauthor{M. Ouali \etal}

\institute{                    
  \inst{1} Recherche Laboratory in Physics and Engineering Sciences, Team of Modern and Applied Physics, FPBM, USMS -  Morocco\\
  \inst{2}  LPFAS, Polydisciplinary Faculty of Safi, UCAM - Morocco}

\abstract{
In the centre of mass frame, we have investigated the process of Higgs production in association with a pair of fermions, $e^{+}e^{-}\rightarrow f\bar{f} H$, at the leading order inside an intense electromagnetic field with circular polarization. Our analytical calculations are based on the Narrow Width Approximation (NWA), which is valid in the leading order. We have considered only the initial particles inside the laser field as a first step. In the second part, we have embedded both initial and final particles in the laser field. We have analyzed the angular distribution of the produced Higgs boson as a function of the laser parameters in both cases. We have found that
in the case where both initial and final particles are embedded in the laser field,
the order of magnitude of the differential cross-section of the process $e^{+}e^{-}\rightarrow\mu^{+}\mu^{-}H$  is reduced more significantly, while that of the process $e^{+}e^{-}\rightarrow\nu\bar{\nu}H$  is enhanced.}
\begin{document}
\maketitle
\section{Introduction}
Studying different aspects of the electromagnetic field effect on physical processes is one of the most important issues of modern and fundamental physics \cite{proc1,proc2,proc3}. This great scientific interest is due to numerous unknown phenomena which are caused by the application of laser radiation, and which provide a deep understanding of the atomic and molecular structure of matter.
These phenomena are of great importance in several fields of physics such as holography, fibreglass optics, biophysics, plasma physics, and nuclear fusion and might be important also in high energy physics. Besides, several authors have been interested in the study of particle physics interactions inside an external field.
In addition, in Refs \cite{dec1,dec2,dec3}, the authors show that the electromagnetic field could affect the physical properties of a particle such as its decay width and lifetime.
In Refs \cite{Z,Hc,Higgs-strahlung,Hn}, we have studied the muon pair production and Higgs pair production inside a laser field with circular polarization, and we have indicated that the total cross-section is affected significantly by the laser field. 

One of the highest priorities of particle physics is to nail down the properties of the Higgs boson \cite{p1,p2,p3,p4,p5,p6}, discovered in 2012 \cite{disc1,disc2}, as precise as possible, and to search for other Higgs bosons if they exist. In addition, for its clean environment, electron-positron colliders provide a good platform to precisely measure various Higgs couplings.
The production of the Higgs boson in association with a pair of fermion, $e^{+}e^{-}\rightarrow f\bar{f} H$, is one of the golden channels to pin down the properties of the Higgs boson in the prospective of $e^{+}e^{-}$ colliders such as CEPC which will function as a Higgs factory in its first run. 
Due to its very clean channel and large cross-section, the $e^{+}e^{-}\rightarrow\mu^{+}\mu^{-}H$ process occupies a special place among numerous Higgs production channels associated with various $Z$-boson decay products to probe the Higgs boson properties. This process is well established in the absence of the electromagnetic field.
For instance, the Leading order contribution to this process is studied in \cite{LO}, while the initial state radiation (ISR) effect is considered in \cite{ISR}. 
In addition, mixed electroweak-QCD corrections to $e^{+}e^{-}\rightarrow\mu^{+}\mu^{-}H$ are studied at CEPC with finite-width effect in \cite{chines}.
Besides, there are enormous works which consider the Higgs-struhlung process $e^{+}e^{-}\rightarrow\nu\bar{\nu}H$ without an external field \cite{nu1,nu2,nu3,nu4}.
In \cite{Higgs-strahlung}, we have analyzed the Higgs-strahlung process at the leading order, and we have found that the circularly polarized laser field reduces its cross-section. 
From the experimental angle, since the $Z$-boson is an unstable particle, it is the decay products of the $Z$-boson rather than the $Z$-boson itself that are tagged by detectors in the Higgs-strahlung production channel. Therefore, to get closer contact with experiment, it is preferable to make precise predictions directly for the process $e^{+}e^{-}\rightarrow f\bar{f} H$, where f represents neutrino or muon. In this respect, we have considered the process $e^{+}e^{-}\rightarrow f\bar{f} H$ in the presence of a plane electromagnetic wave with circular polarization to analyze its dependence on the laser parameters. Moreover, we have compared the angular distribution of the Higgs boson inside a laser field for both processes.

The remainder of this paper is structured as follows: In the next section section, we have performed a theoretical calculation of the differential cross-section of the process $e^{+}e^{-}\rightarrow f\bar{f} H$ inside an electromagnetic field by using the narrow width approximation. In the first part, we have only embedded the initial $e^{+}e^{-}$ beam in the laser field. Then, we have performed, in the second part, a calculation in which we have considered both the initial and final particles inside the laser field. The third section is devoted to the numerical analysis of the obtained results. In the forth section, we give a summary of the obtained results.
\section{Outline of the theory}\label{sec:theory} 
In this part, we will give an overview of the analytical calculation of the differential cross-section of the process $e^{+}e^{-}\rightarrow f\bar{f} H$ in the presence of a circularly polarized electromagnetic field. The latter is supposed, at the first step, to be interacted with the initial particles, while the three final particles are free. As we can see from figure \ref{fig1}, it is possible for the pair of fermion-antifermion to be produced at the resonance from the on-shell $Z$-boson. In addition, since $\Gamma_{Z}<<M_{Z}$ and the production rate is predominant at the $Z$-boson resonance, the Narrow Width Approximation (NWA) \cite{NWA} might be a good approximation to calculate the production cross-section. 
The processes considered, here, is denoted as:
\small
\begin{equation}
e^{-}(q_{1}, s_{1})\,e^{+}(q_{2}, s_{2})\rightarrow f(q_{3},s_{3})\,\bar{f}(q_{4},s_{4})H(p_{H})\,;\, f=\left\{\mu,\nu \right\}\nonumber,
\end{equation}
\normalsize
where the momenta of the incoming and outgoing particles are specified in the parentheses. $s_{i}(i=1,2,3,4)$ denote their spins.
These processes can be described  at the lowest order by the s-channel Feynman diagrams as depicted in figure \ref{fig1}:
\begin{figure*}[t!]
  \centering
      \includegraphics[scale=0.35]{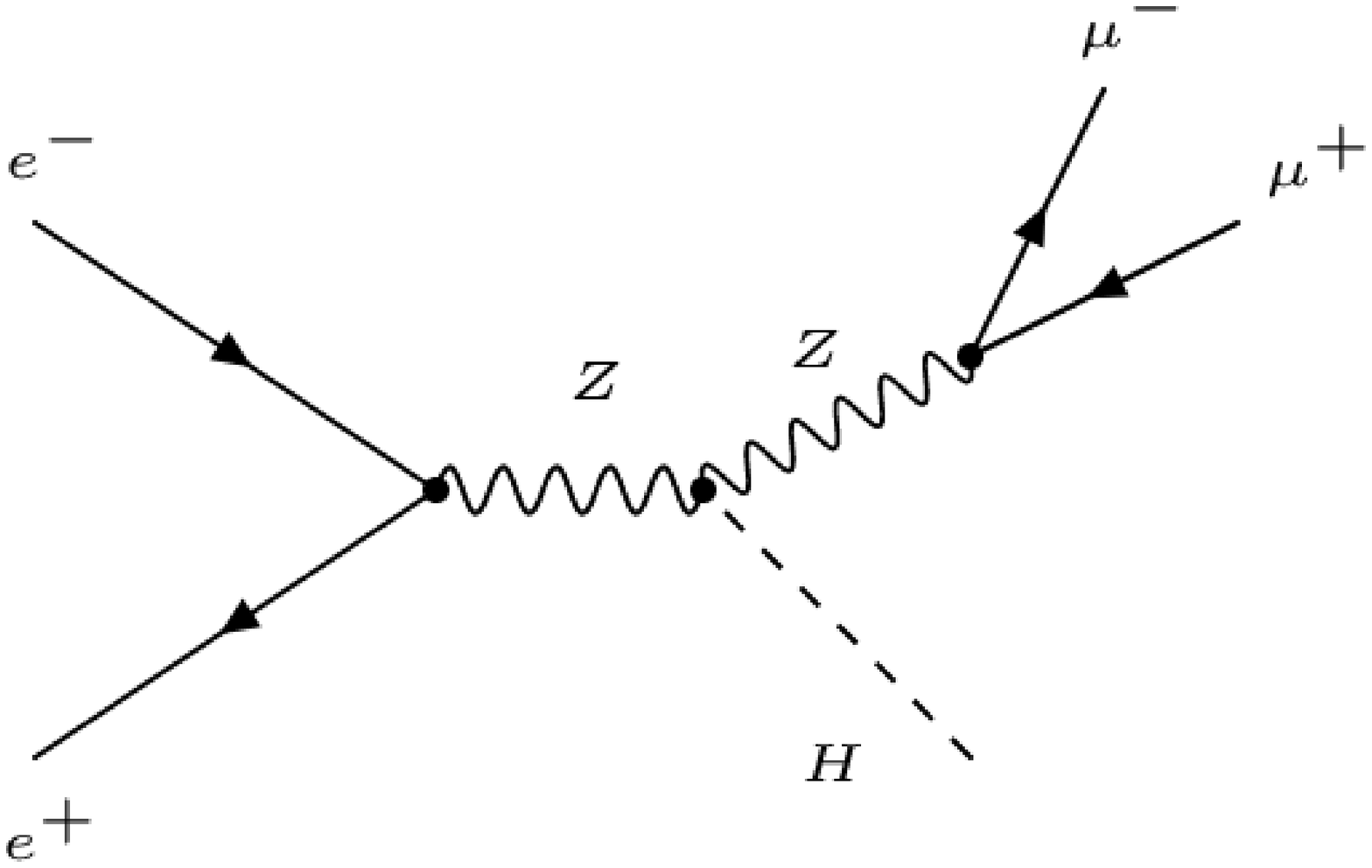}\hspace*{0.6cm}
      \includegraphics[scale=0.35]{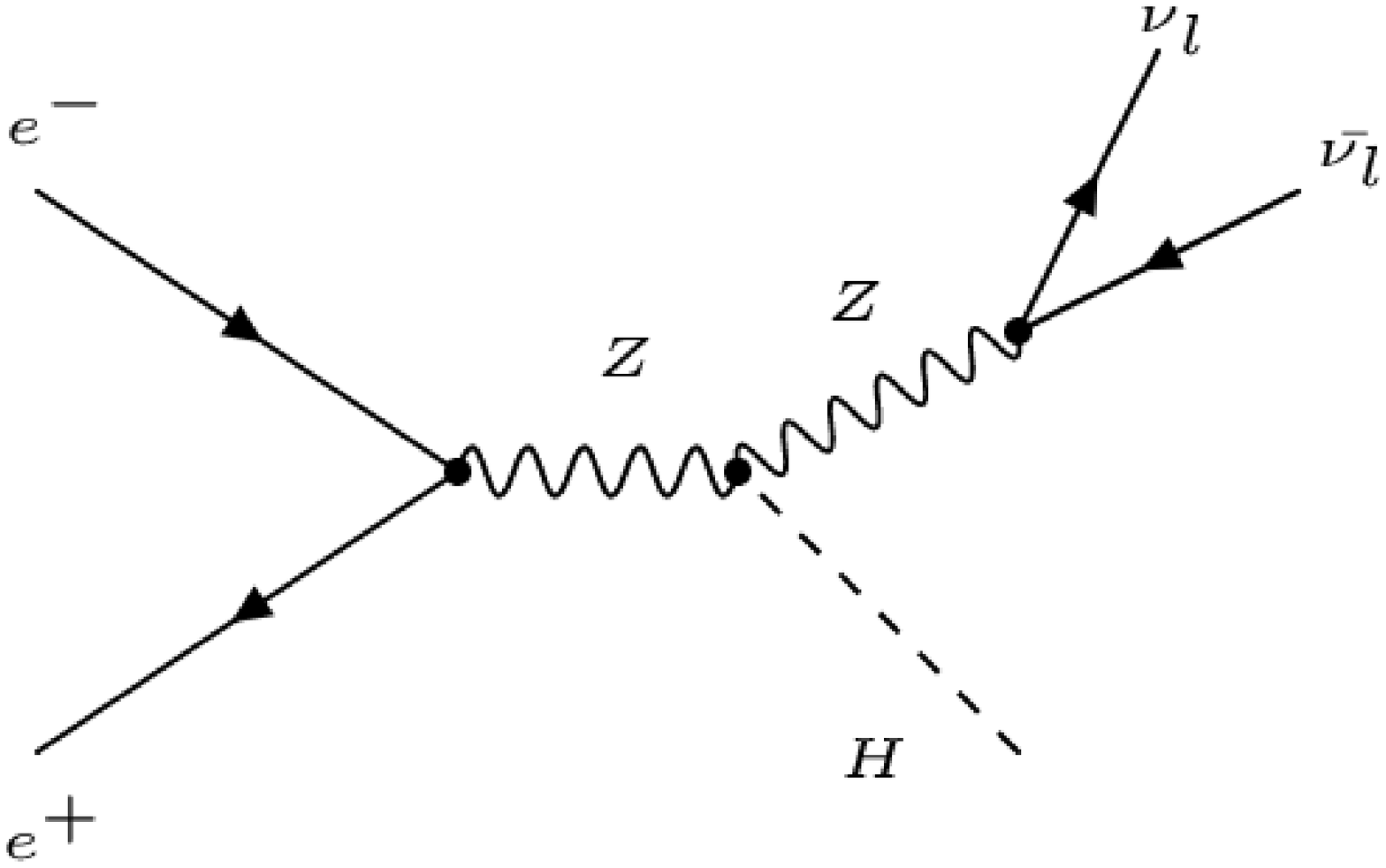}
        \caption{Leading order s-channel Feynman diagrams for the processes: $e^{+}e^{-}\rightarrow\mu^{+}\mu^{-}H$ (left);  $e^{+}e^{-}\rightarrow\nu\bar{\nu}H$ (right).}
        \label{fig1}
\end{figure*}
\subsection{Embedding the initial particles in an electromagnetic field with circular polarization}
By using Narrow Width Approximation (NWA), the differential cross section of muon/ neutrion pair production in association with a Higgs boson can be expressed by:
\begin{eqnarray}
\dfrac{d\sigma_{n_{(e^{+}e^{-}\longrightarrow f\bar{f} H)}}}{d\cos\theta_{H}}=\dfrac{d\sigma_{n_{(e^{+}e^{-}\longrightarrow ZH)}}}{d\cos\theta} Br_{0_{(Z\rightarrow f\bar{f})}},
\label{1}
\end{eqnarray}
where $\theta_{H}$ is the angle between the incoming electron and the produced Higgs boson in the centre of mass frame. $Br_{0_{(Z\rightarrow f\bar{f})}}$ represents the branching ratio of $Z$-boson decay to a pair of muons or neutrinos.
The quantity  $\dfrac{d\sigma_{n_{(e^{+}e^{-}\longrightarrow ZH)}}}{d\cos\theta}$ denotes the differential cross-section of the Higgs-strahlung production in the presence of a circularly polarized laser field, and it is expressed as follows:
\begin{eqnarray}
\dfrac{d\sigma_{n_{(e^{+}e^{-}\longrightarrow ZH)}}}{d\cos\theta}&=&\nonumber\dfrac{2\pi\, g^{2}M_{Z}^{4}}{64\cos_{\theta_{W}}^{2}v^{2}} \big|\overline{M_{fi}^{n}} \big|^{2}  \dfrac{2|\mathbf{p_{H}}|^{2}}{(2\pi)^{2}Q_{H}} \\&\times & \Bigg(  \dfrac{1}{(q_{1}+q_{2}+nk)^{2}-M_{Z}^{2}} \Bigg)^{2} \\&\times & \nonumber \dfrac{1}{\sqrt{(q_{1}q_{2})^{2}-m_{e}^{*^{4}}}} \dfrac{1}{\big|g^{'}(|\mathbf{p}_{H}|)\big|_{g(|\mathbf{p}_{H}|)=0}},  
\label{2}
\end{eqnarray}
where $n$ is interpreted as the number of transferred photons between the electromagnetic field and the $e^{+}e^{-}$ initial beam. $\theta_{W}$ is the Weinberg angle, and $g$ is the electroweak coupling constant. {$ v=(\sqrt{2}G_{F})^{-\frac{1}{2}} $} is the vacuum expectation value with $G_{F}=(1.16637\pm 0.00002)\times 10^{-11}\,MeV^{-2}$ is the Fermi coupling constant. $M_{Z}$ is the mass of the weak $Z$-boson. $\mathbf{ p_{H}}$ is the four-momentum of the produced Higgs boson, and $Q_{H}$ denotes its effective energy. $m_{e}^{*}$ is the effective mass of the electron and positron inside the electromagnetic field. 
The spinorial part, $\big|\overline{M_{fi}^{n}} \big|^{2}$, can be expressed as a product of traces in the Dirac algebra by summing over the final polarizations and averaging over the initial ones.  
The function $g^{'}(|\mathbf{p}_{H}|)$ is expressed as follows:
\begin{equation}
 g^{'}(|\mathbf{p}_{H}|)=\dfrac{-4 e^{2}a^{2}}{\sqrt{s}}\dfrac{|\mathbf{p}_{H}|}{\sqrt{|\mathbf{p}_{H}|^{2}+M_{H}^{2}}}-\dfrac{2|\mathbf{p}_{H}|(\sqrt{s}+n\omega)}{\sqrt{|\mathbf{p}_{H}|^{2}+M_{H}^{2}}},
 \label{3}
\end{equation}
where $M_{H}$ denotes the Higgs boson mass. $\sqrt{s}$ is the centre of mass energy of the collider, and $\omega$ is the frequency of the laser field. $a$ represents the amplitude of the electromagnetic field which is given by its classical four potential $A^{\mu}(\phi)=a_{1}^{\mu}\cos\phi+a_{2}^{\mu}\sin\phi$ such that $a_{1}^{2}=a_{2}^{2}=a^{2}=-|\mathbf{a}|^{2}=-\big(\varepsilon/\omega \big)^{2}$ where $\phi=(k.x)$ is the laser field phase and $\varepsilon$ its strength. The electromagnetic field is considered to be propagating along the $z$-axis, while $e^{+}$ and $e^{-}$  are chosen in $x$ and $-x$ directions, respectively, in the centre of mass frame.
The branching ratio of $Z$-boson decay into a pair of fermions is expressed as follows:
\begin{equation}
Br_{0_{(Z\rightarrow f\bar{f})}}=\dfrac{\Gamma_{0_{(Z\rightarrow f\bar{f} )}}}{\Gamma_{Z}},
\label{4}
\end{equation}
where $\Gamma_{Z}=(2.4952\pm 0.0023)\,GeV$ is the total decay width of the weak $Z$-boson. $\Gamma_{0_{(Z\rightarrow f\bar{f} )}}$ denotes its partial decay width to a pair of muons or neutrinos, and its  expression is given by \cite{Thomson}:
\begin{equation}
\Gamma_{0_{(Z\rightarrow f\bar{f})}}=\dfrac{g^{4}}{\cos^{4}(\theta_{W})}\dfrac{M_{Z}}{48\pi}\Big[ g_{v}^{2}+ g_{a}^{2} \Big],
\label{5} 
\end{equation}
where $g_{v}$ and $g_{a}$ are successively the vector and axial vector coupling constant
such that $g_{v}=-1+4\sin^{2}(\theta_{W})$ and $g_{a}=1$ for muons and $g_{v}=g_{a}=1$ for neutrinos.
\subsection{Dressing both initial and final particles}
We proceed, in this part, in the same manner as in the previous one by dressing both the incoming and outgoing particles that contribute to the studied process. Since the produced Higgs boson is a neutral massive particle, then it will not interact with the electromagnetic field. Therefore, it can be described by Klein-Gordon states while the fermions are described by Dirac-Volkov states \cite{Volkov}. The angular distribution of the cross-section by using the NWA can be expressed as follows:
\begin{equation}
\dfrac{d\sigma_{n,N_{(e^{+}e^{-}\longrightarrow f\bar{f} H)}}}{d\cos\theta_{H}}=\dfrac{d\sigma_{n_{(e^{+}e^{-}\longrightarrow ZH)}}}{d\cos\theta}Br_{N_{(Z\rightarrow f\bar{f} )}},
\label{6}
\end{equation}
where $n$ is the number of photons that is exchanged between the electromagnetic field and the initial beam, while $N$ is that exchanged with the final particles. The laser-assisted branching ratio of $Z$-boson invisible decay can be expressed as follows:
\begin{equation}
Br_{N_{(Z\rightarrow\nu_{l}\bar{\nu_{l}})}}=\dfrac{\Gamma_{0_{(Z\rightarrow\nu_{l}\bar{\nu_{l}})}}}{\Gamma^{N}_{tot}},
\label{7}
\end{equation}
where $\Gamma_{0_{(Z\rightarrow\nu_{l}\bar{\nu_{l}})}}$ is the $Z$-boson laser-assisted partial decay width, which is the same as its corresponding laser free-partial decay width due to the fact that neutrinos are neutral particles. $\Gamma^{N}_{tot}$ is the total decay width inside the electromagnetic field, and it can be decomposed as follows:
\begin{eqnarray}
\Gamma^{N}_{tot}&=&\nonumber\Gamma^{N}_{(Z\rightarrow hadrons)}+\Gamma_{({Z\rightarrow \nu \bar{\nu}})}+\Gamma^{N}_{{({Z\rightarrow \mu \bar{\mu}})}}\\&+&\Gamma^{N}_{({Z\rightarrow other\, leptons})}.
\label{8}
\end{eqnarray}
Since muons are charged particles, they will interact with the laser field. Therefore, $\Gamma^{N}_{tot}$ will be different to  the $Z$-boson total decay width in the absence of the electromagnetic field.
The branching ratio of the muonic decay mode of $Z$-boson in the presence of a circularly polarized laser field is expressed as follows:
\begin{equation}
Br_{N_{(Z\rightarrow\mu^{+}\mu^{-})}}=\dfrac{\Gamma_{N_{(Z\rightarrow\mu^{+}\mu^{-})}}}{\Gamma^{N}_{tot}},
\label{9}
\end{equation}
where the partial decay width $\Gamma_{N_{(Z\rightarrow\mu^{+}\mu^{-})}}$ is given by the following expression:
\begin{equation}
\Gamma_{N_{(Z\rightarrow\mu^{+}\mu^{-})}}=\dfrac{G_{F}M_{Z}N_{c}}{16\sqrt{2}(2\pi)^{2}}\int\dfrac{|\mathbf{q_{3}}|^{2}d\Omega_{3}}{Q_{3}Q_{4}{g ^{\prime}}(|\mathbf{q_{3}}|)}\big| \overline{M_{fi}^{N}} \big|^{2},
\label{10}
\end{equation}
where $N_{c}$ indicates the number of colors such that $N_{c}=3$ for quarks, and $N_{c}=1$ for the other categories.
$\mathbf{q_{3}}$ is the effective four-momentum of the produced fermion. $Q_{3}$ and $Q_{4}$ are the energies acquired by the fermion and its corresponding antifermion, respectively, inside the external field.
with $g^{ \prime}(|\mathbf{q_{3}}|)$ is expressed as follows:
 \begin{eqnarray}
g^{ \prime}(|\mathbf{q_{3}}|)&=&\nonumber \dfrac{|\mathbf{q_{3}}|-N\omega\cos(\theta)}{\sqrt{(N\omega)^{2}+|\mathbf{q_{3}}|^{2}-2N\omega|\mathbf{q_{3}}|\cos(\theta)+m_{f}^{*^{2}}}}\\&+&\dfrac{|\mathbf{q_{3}}|}{\sqrt{|\mathbf{q_{3}}|^{2}+m_{f}^{*^{2}}}}.
\label{11}
\end{eqnarray}
where $m_{f}^{*}$ is the effective mass of the muon.
\section{Results and discussion}\label{section-results}
In this section, we present and discuss the obtained numerical results about the angular distribution of the Higgs boson production in association with a pair of muons or neutrinos in the centre of mass frame. The $e^{+}e^{-}$ initial beam is submitted in a strong laser field with circular polarization. The centre of mass-energy is chosen as $\sqrt{s}=240\, GeV$ at which the future $e^{+}e^{-}$ colliders will operate as Higgs factories. 
The standard model parameters are taken from PDG \cite{PDG} such that: The electron mass $m_{e} = 0.511\, MeV$, the muon mass $m_{\mu} = 105.66\, MeV$, the electroweak mixing angle $\sin^{2}(\theta_{W})=0.23126$, the mass of the $Z$-boson $M_{Z}=91.186\,GeV$ and its total decay width $\Gamma_{Z}=2.4952\,GeV$. The produced neutrinos are considered as massless particles. 
\begin{figure*}[t!]
  \centering
      \includegraphics[scale=0.58]{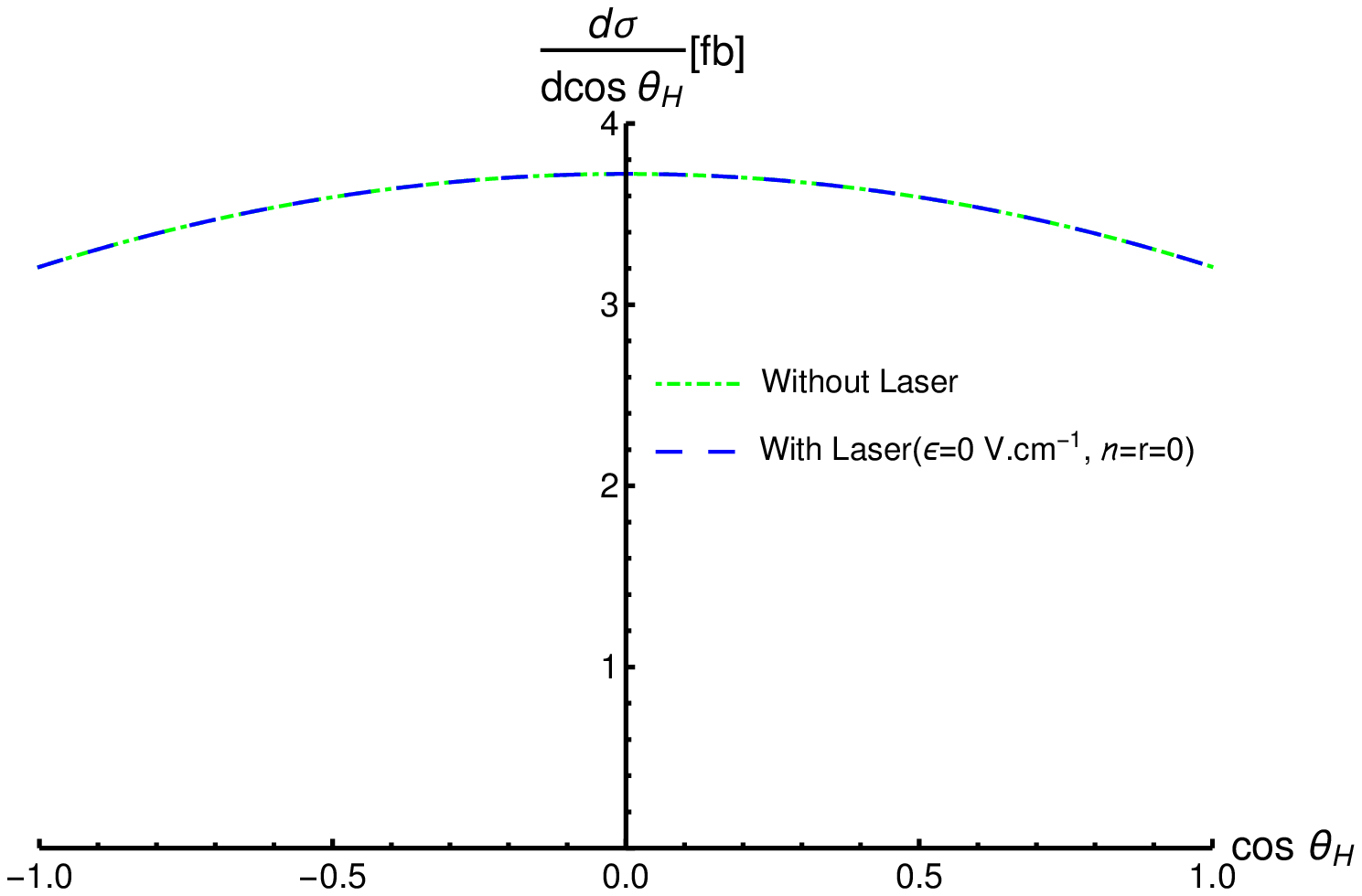}\hfill
      \includegraphics[scale=0.58]{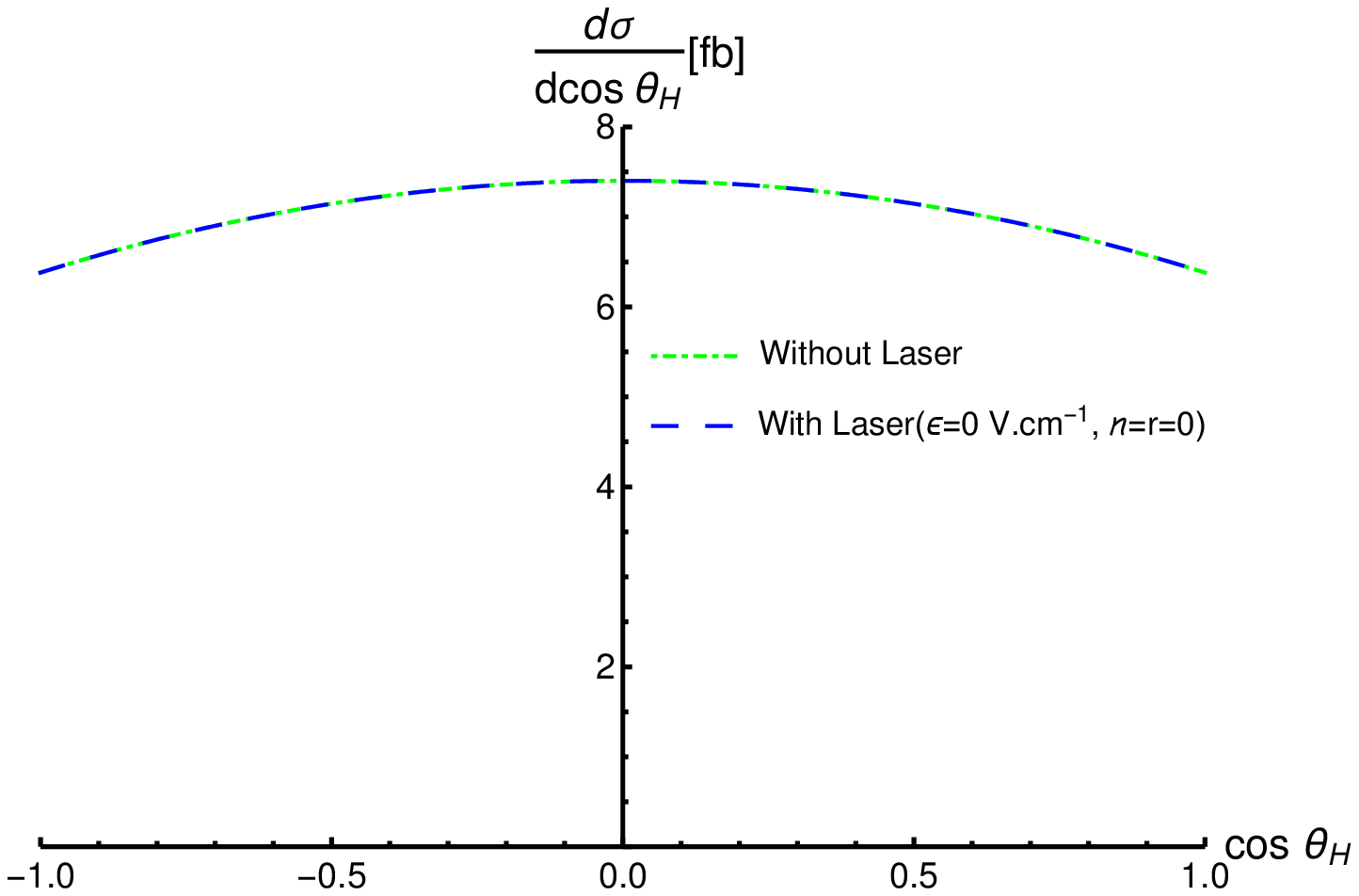}\par\medskip
        \caption{Comparison between the angular distribution of the cross section in the absence and presence of the laser field for both $e^{+}e^{-}\rightarrow\mu^{+}\mu^{-}H$ (left) and $e^{+}e^{-}\rightarrow\nu\bar{\nu}H$ (right) by considering $\sqrt{s}=240\,GeV$.}
        \label{fig2}
\end{figure*}
To check the validity of the NWA for this process inside the laser field, we have compared, in figure \ref{fig2}, the laser-free angular distribution of the Higgs boson production in association with whether a pair of muons (left panel) or neutrino (right panel) with that corresponds to the presence of an external field. In addition, the angular distribution of the Higgs boson in the absence of the laser field is calculated by using the scattering matrix approach while the laser-assisted angular distribution is obtained by using the NWA. The laser parameters are chosen as: $\varepsilon=0\,V.cm^{-1}$ and $n=0$.
From this figure, it is obvious that the two cross-sections are in good agreement for all values of $\cos (\theta_{H})$. Therefore, this comparison supports not only the validity of NWA approximation but also the validity of our calculations in the presence of an electromagnetic field.
To study the behavior of the partial differential cross section inside the electromagnetic field, we have presented in tables (\ref{tab1}) and (\ref{tab2}) the differential cross section as a function of the number of exchanged photons for both $ e^{+}e^{-}\rightarrow \mu^{+}\mu^{-}H $ and $ e^{+}e^{-}\rightarrow \nu\bar{\nu}H $, respectively. We have chosen two different known laser sources which are \textbf{Nd:YAG laser} and \textbf{He:Ne laser} with three laser field strengths $\varepsilon=10^{6}\,V.cm^{-1}$, $\varepsilon=10^{7}\,V.cm^{-1} $ and $\varepsilon=10^{8}\,V.cm^{-1}$. We mention that the negative values correspond to the emission of photons while the positive values correspond to the absorption.
\begin{table*}[t!]
\addtolength{\tabcolsep}{-4pt}
\centering
\caption{\label{tab1}Laser-assisted differential cross section of $ e^{+}e^{-}\rightarrow \mu^{+}\mu^{-}H  $ as a function of the number of exchanged photons for two different known laser sources by taking  $\sqrt{s}=240\,GeV$ and $\cos(\theta_{H})=1$.}
\begin{tabular}{ccccccccc}
\hline
  & & $ d\sigma_{n}/d\cos \theta_{H} $[fb]  & &$ d\sigma_{n}/d\cos \theta_{H} $[fb] & & $ d\sigma_{n}/d\cos \theta_{H} $[fb] &  \\
 Laser's type  & n &  $\varepsilon=10^{6}(\,V.cm^{-1})  $ & n &$\varepsilon=10^{7}(\,V.cm^{-1}) $& n & $\varepsilon=10^{8}(\,V.cm^{-1}) $ & \\
 \hline
 &$\pm150$ & $ 0 $ & $\pm1400$ & $ 0 $ & $\pm15000$ & $ 0 $ \\
   &$\pm120$ & $ 0 $ & $\pm1100$ & $ 0 $ & $\pm12000$ & $ 0 $ \\
  \textbf{Nd:YAG laser}  &$\pm90$ & $ 4.25272\times 10^{-2} $ & $\pm900$ & $ 4.18579\times 10^{-3} $ & $\pm9000$ & $ 1.20792\times 10^{-4} $ \\
  $\omega=1.17\,(eV) $ &$\pm60$ & $ 2.49174\times 10^{-2} $ & $\pm600$ & $ 2.50017\times 10^{-3} $ & $\pm6000$ & $ 1.8783\times 10^{-5} $ \\
    &$\pm30$ & $ 1.45116\times 10^{-3} $ & $\pm300$ & $ 1.92703\times 10^{-3} $ & $\pm3000$ & $ 1.22777\times 10^{-4} $ \\
    &$0$ & $ 1.51472\times 10^{-2}  $ & $0$ & $ 1.92442\times 10^{-3} $  & $0$ & $ 2.25656\times 10^{-6} $ \\
     \hline
 &$\pm50$ & $ 0 $ & $\pm500$ & $ 0 $ & $\pm4800$ & $ 0 $ \\
   &$\pm40$ & $ 0 $ & $\pm400$ & $ 0 $ & $\pm3800$ & $ 0 $ \\
   \textbf{He:Ne laser} &$\pm30$ & $ 4.43794\times 10^{-2} $ & $\pm300$ & $ 6.80208\times 10^{-3} $ & $\pm3000$ & $ 1.07927\times 10^{-6} $ \\
  $\omega=2\,(eV) $ &$\pm20$ & $ 3.02771\times 10^{-2} $ & $\pm200$ & $ 2.04909\times 10^{-5} $ & $\pm2000$ & $ 4.24357\times 10^{-5} $ \\
   &$\pm10$ & $ 2.08478\times 10^{-2} $ & $\pm100$ & $ 2.31443\times 10^{-3} $ & $\pm1000$ & $ 4.82931\times 10^{-4} $ \\
    &$0$ & $ 4.59189\times 10^{-2}  $ & $0$ & $ 1.90147\times 10^{-3} $  & $0$ & $ 4.39443\times 10^{-4} $ \\
     \hline
\end{tabular}
\end{table*}
\begin{table*}[t!]
\addtolength{\tabcolsep}{-4pt}
\centering
\caption{\label{tab2}Laser-assisted differential cross section of $ e^{+}e^{-}\rightarrow \nu\bar{\nu}H $ as a function of the number of exchanged photons for two different known laser sources by taking $\sqrt{s}=240\,GeV$ and $\cos(\theta_{H})=1$.}
\begin{tabular}{ccccccccc}
\hline
  & & $ d\sigma_{n}/d\cos \theta_{H} $[fb]  & &$ d\sigma_{n}/d\cos \theta_{H} $[fb] & & $ d\sigma_{n}/d\cos \theta_{H} $[fb] &  \\
 Laser's type & n &  $\varepsilon=10^{6}(\,V.cm^{-1})  $ & n &$\varepsilon=10^{7}(\,V.cm^{-1}) $& n & $\varepsilon=10^{8}(\,V.cm^{-1}) $ & \\
 \hline
 &$\pm150$ & $ 0 $ & $\pm1400$ & $ 0 $ & $\pm15000$ & $ 0 $ \\
   &$\pm120$ & $ 0 $ & $\pm1100$ & $ 0 $ & $\pm12000$ & $ 0 $ \\
    \textbf{Nd:YAG laser} &$\pm90$ & $ 8.45773\times 10^{-2} $ & $\pm900$ & $ 8.32463\times 10^{-3} $ & $\pm9000$ & $ 2.40228\times 10^{-4} $ \\
  $\omega=1.17\,(eV) $ &$\pm60$ & $ 4.95552\times 10^{-2} $ & $\pm600$ & $ 4.97229\times 10^{-3} $ & $\pm6000$ & $ 3.73554\times 10^{-5} $ \\
    &$\pm30$ & $ 2.88604\times 10^{-3} $ & $\pm300$ & $ 3.83243\times 10^{-3} $ & $\pm3000$ & $ 2.44177\times 10^{-4} $ \\
    &$0$ & $ 3.01245\times 10^{-2}  $ & $0$ & $ 3.82725\times 10^{-3} $  & $0$ & $ 4.48781\times 10^{-6} $ \\
     \hline
 &$\pm50$ & $ 0 $ & $\pm500$ & $ 0 $ & $\pm4800$ & $ 0 $ \\
   &$\pm40$ & $ 0 $ & $\pm400$ & $ 0 $ & $\pm3800$ & $ 0 $ \\
  \textbf{He:Ne laser} &$\pm30$ & $ 8.82609\times 10^{-2} $ & $\pm300$ & $ 1.35278\times 10^{-2} $ & $\pm3000$ & $ 2.14644\times 10^{-6} $ \\
  $\omega=2\,(eV) $  &$\pm20$ & $ 6.02146\times 10^{-2} $ & $\pm200$ & $ 4.0752\times 10^{-5} $ & $\pm2000$ & $ 8.43953\times 10^{-5} $ \\
   &$\pm10$ & $ 4.14616\times 10^{-2} $ & $\pm100$ & $ 4.60289\times 10^{-3} $ & $\pm1000$ & $ 9.60444\times 10^{-4} $ \\
    &$0$ & $ 9.13227\times 10^{-2}  $ & $0$ & $ 3.7816\times 10^{-3} $  & $0$ & $ 8.73956\times 10^{-4} $ \\
     \hline
\end{tabular}
\end{table*}
In addition, for the two processes and regardless of the laser source, the differential cross section in case of emission of $n$ photons is equal to that corresponds to absorption of $n$ photons. This symmetry is mainly due to the presence of Bessel functions which are characterized by the fact that $J_{n}(z)=(-1)^{n}J_{-n}(z)$. From the tables (\ref{tab1}) and (\ref{tab2}), we remark that the partial differential cross section for each process becomes null whenever the number of transferred photons overcomes a certain threshold value. The latter is called cut-off, and it depends on the laser frequency and its strength. In addition, it increases as far as the frequency decreases or by increasing the laser field strength. For instance, for the process $ e^{+}e^{-}\rightarrow \nu\bar{\nu}H $ and for $\omega=1.17\,(eV), $ the cutoff is $n=\pm120$, $n=\pm1100$ and $n=\pm12000$ for $\varepsilon=10^{6}(\,V.cm^{-1}) $, $\varepsilon=10^{7}(\,V.cm^{-1}) $ and $\varepsilon=10^{8}(\,V.cm^{-1}) $, respectively. By comparing results of table (\ref{tab1}) and table (\ref{tab2}), we observe that the cutoffs that correspond to a given laser field strength and frequency are the same for both processes $ e^{+}e^{-}\rightarrow \mu^{-} \mu^{+}H $ and $ e^{+}e^{-}\rightarrow \nu\bar{\nu}H $. Therefore, the cutoff depends only on the laser parameters. To understand the effect of the electromagnetic field on the summed differential cross section, we have plotted in figure (\ref{fig3}) the behavior of the angular distribution of the Higgs boson for different number of transferred photons.
\begin{figure*}[t!]
  \centering
      \includegraphics[scale=0.58]{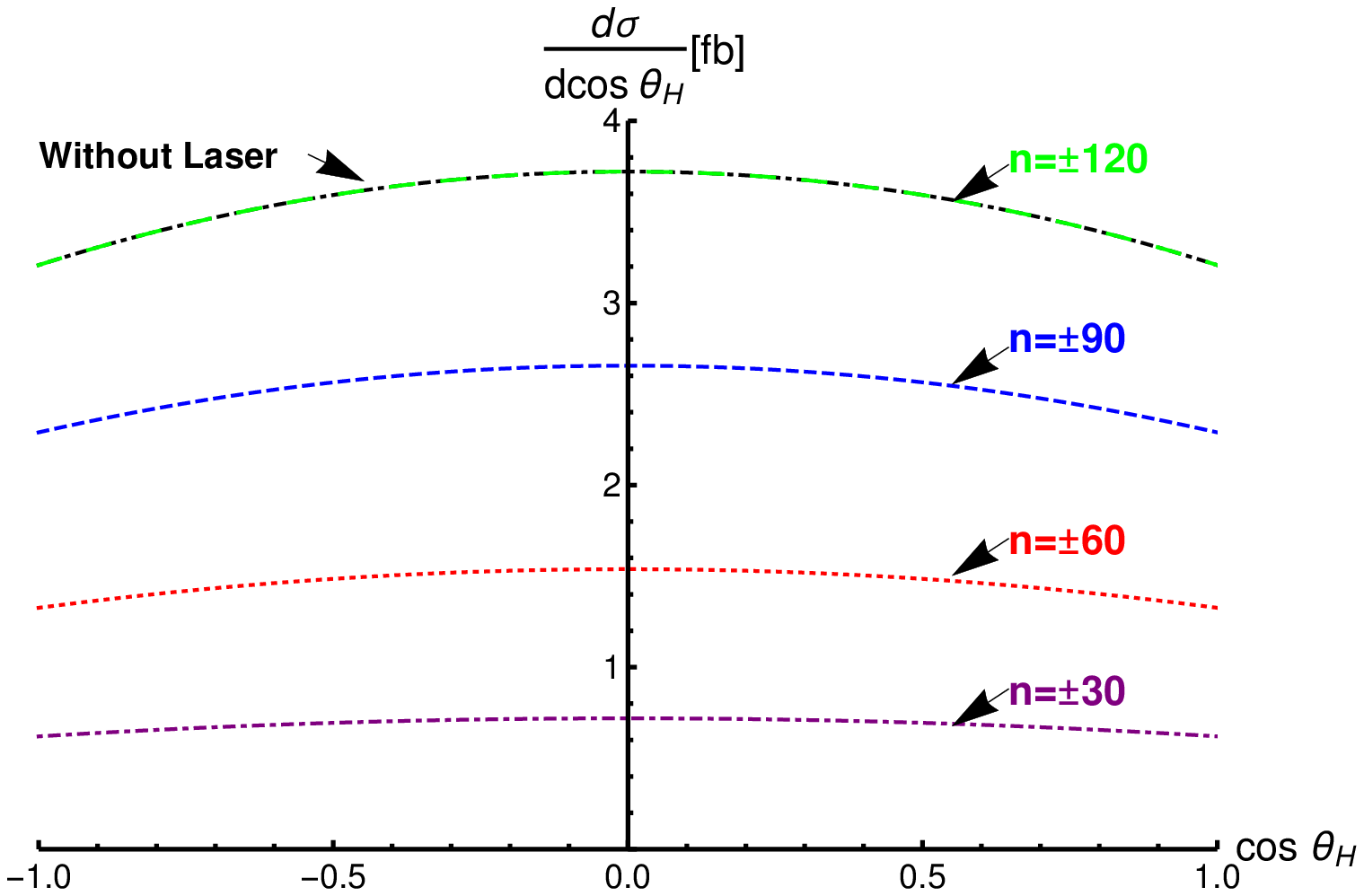}\hfill
      \includegraphics[scale=0.6]{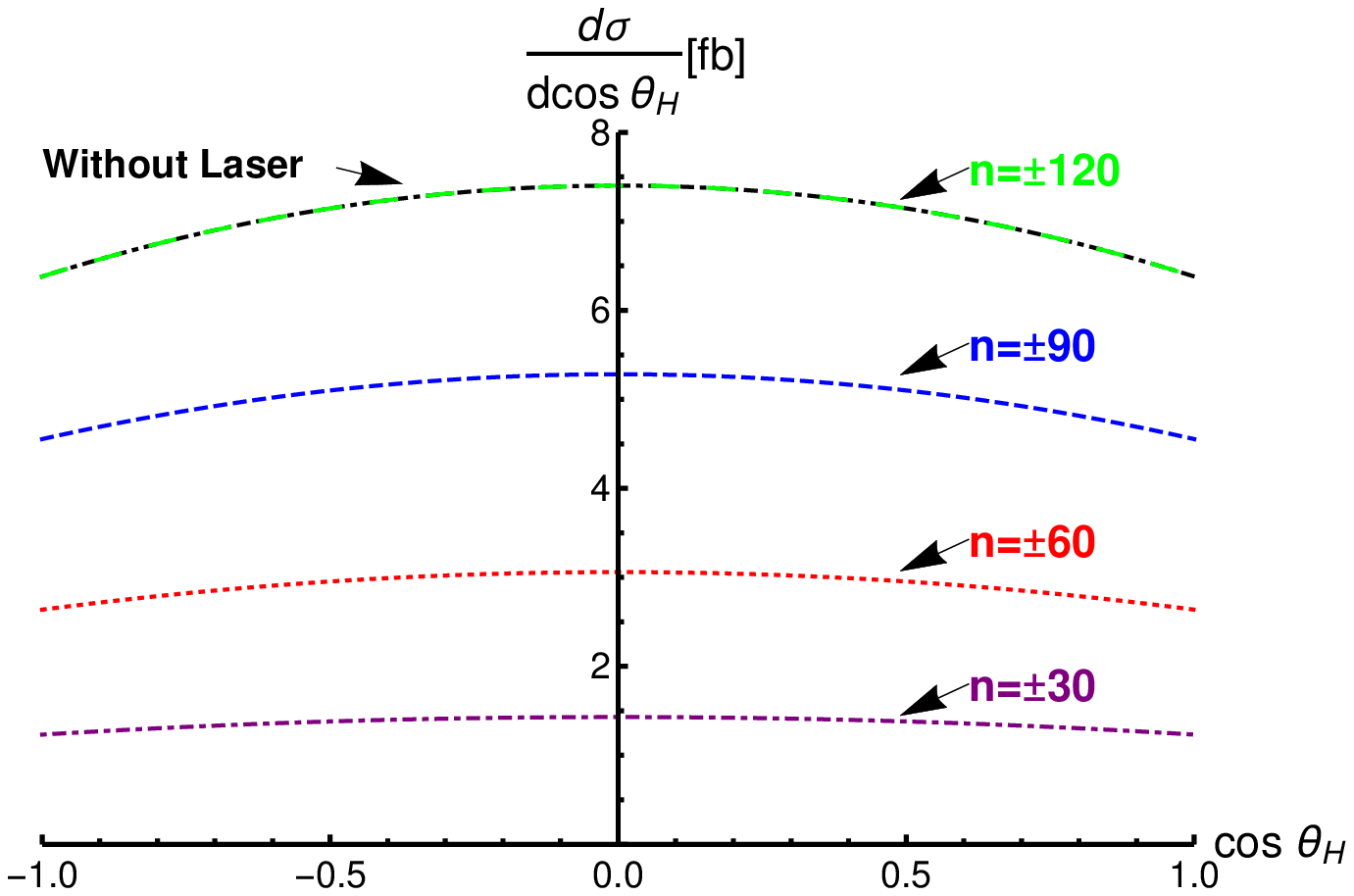}\par\medskip
        \caption{Angular distribution of the Higgs boson inside the laser field at $\sqrt{s}=240\,GeV$ for various number of exchanged photons for both $e^{+}e^{-}\rightarrow\mu^{+}\mu^{-}H$ (left) and $e^{+}e^{-}\rightarrow\nu\bar{\nu}H$ (right). We have chosen: $\varepsilon=10^{6}\, V.cm^{-1}$, $\omega=1.17\,eV$ and $\sqrt{s}=240\,GeV$.}
        \label{fig3}
\end{figure*}
Figure \ref{fig3} represents the number of exchanged photons dependence of the angular distribution of the produced Higgs boson, which is associated with a pair of muons (left panel) or a pair of neutrinos (right panel). We can see from this figure that for both processes and both the absence and presence of a laser field, the differential cross-section is maximum for $\cos(\theta_{H})=0$, and it decreases symmetrically with respect to $\cos(\theta_{H})=0$. We remark that the laser-assisted differential cross-section of both processes decreases inside the electromagnetic field, and it depends on the summation over the number of exchanged photons. In addition, by summing over $n$ from $-120$ to $+120$, the cross-section becomes equal to its corresponding laser-free cross-section, and this summation is called sum-rule \cite{Watson1,Watson2}. According to the table \ref{tab1} and for $\omega=1.17\,eV$, $n=\pm120$ corresponds to the cutoff at which the partial cross-section vanishes. Therefore, there will be no exchange of photons between the initial $e^{+}e^{-}$ beam and the laser beam. However, the differential cross-section is reduced inside the electromagnetic field in the case where the summation is done over a number $n$ inferior to the corresponding cutoff. In addition, this effect decreases by increasing the number of exchanged photons. Another important point to be mentioned is that the order of magnitude of the cross-section is different from one figure to another. For instance, for the same number of exchanged photons $n=\pm 90$, the maximum differential cross-section is approximately $3.8\,fb$ and $7.4\,fb$ for the two processes $e^{+}e^{-}\rightarrow\mu^{+}\mu^{-}H$ and $e^{+}e^{-}\rightarrow\nu\bar{\nu}H$, respectively.
\begin{table*}[t!]
\small
\addtolength{\tabcolsep}{-4pt}
\centering
\caption{\label{tab3}Angular distribution of the Higgs boson as a function of the laser field strength for both $e^{+}e^{-}\rightarrow\mu^{+}\mu^{-}H$ (columns 3 and 4) and $e^{+}e^{-}\rightarrow\nu\bar{\nu}H$ (columns 5 and 6) by taking $n=N=\pm10$, $\sqrt{s}=240\,GeV$ and $\cos(\theta_{H})=1$.}
\begin{tabular}{ccccccccc}
\hline 
&  \multirow{1}{*}{}&\multicolumn{2}{c}{$ e^{+}e^{-}\rightarrow \mu^{+}\mu^{-}H  $}& \multicolumn{2}{c}{$ e^{+}e^{-}\rightarrow \nu\bar{\nu}H  $}\\
\cmidrule(lr){3-4} \cmidrule(lr){5-6} Laser's type  & $\varepsilon(\,V.cm^{-1})  $& $ d\sigma_{n}/d\cos \theta_{H} $[fb] &$ d\sigma_{n,N}/d\cos \theta_{H} $[fb] & $ d\sigma_{n}/d\cos \theta_{H} $[fb] &$ d\sigma_{n,N}/d\cos \theta_{H} $[fb] &  \\
\hline
 &$10$ & $ 3.20842 $ & $ 3.20842 $ & $ 6.38084 $ & $ 6.38084 $ \\
   &\vdots &  \vdots   & \vdots &  \vdots   & \vdots  \\
     &$10^{4}$ & $ 3.20842 $ & $ 3.20842 $ &$ 6.38084 $ & $ 6.38084 $ \\
   &$10^{5}$ & $ 3.07595 $ & $ 3.07595 $  &   $ 6.11739 $   &   $ 6.1174 $    \\
    &$10^{6}$ & $ 0.207585 $ & $ 6.39261\times 10^{-2} $ &   $ 0.412841 $   &  $ 0.61267 $  \\
    \textbf{Nd:YAG laser} &$10^{7}$ & $ 2.22209\times 10^{-2}  $ & $ 1.78754\times 10^{-3} $ & $ 4.41926\times 10^{-2}  $ & $ 0.181988 $  \\
    $\omega=1.17\,(eV) $ &$10^{8}$ & $ 2.02788\times 10^{-3}  $ & $ 1.87977\times 10^{-5} $ & $ 4.033\times 10^{-3}  $ & $ 1.92697\times 10^{-2}  $  \\
    &$10^{9}$ & $ 2.04008\times 10^{-4}  $ & $ 1.96667\times 10^{-7} $ &  $ 4.05728\times 10^{-4}  $ &  $ 1.96849\times 10^{-3}  $ \\
    &$10^{10}$ & $ 2.20197\times 10^{-5}  $ & $ 2.07209\times 10^{-9} $  & $ 4.37924\times 10^{-5}  $ & $ 2.12807\times 10^{-4}  $ \\
    &$10^{11}$ & $ 2.03945\times 10^{-6}  $ & $ 1.99601\times 10^{-11} $ & $ 4.05602\times 10^{-6}  $ & $ 1.97131\times 10^{-5}  $  \\
    &$10^{12}$ & $ 2.04431\times 10^{-7}  $ & $ 1.91834\times 10^{-13} $ & $ 4.06568\times 10^{-7}  $ & $ 1.97604\times 10^{-6}  $ \\
    &$10^{13}$ & $ 2.11952\times 10^{-8}  $ & $ 2.10812\times 10^{-15} $  & $ 4.21527\times 10^{-8}  $ &  $ 2.04875\times 10^{-7}  $ \\
     \hline
     &$10$ & $ 3.20842 $ & $ 3.20842 $ & $ 6.38084 $ & $ 6.38084 $ \\
   &\vdots &  \vdots   & \vdots &  \vdots   & \vdots   \\
     &$10^{5}$ & $ 3.20842 $ & $ 3.20842 $ &$ 6.38084 $ & $ 6.38084 $ \\
    &$10^{6}$ & $ 0.612254 $ & $ 0.501391 $ & $ 1.21764 $ &  $ 1.25673 $  \\
    &$10^{7}$ & $ 6.20036\times 10^{-2}  $ & $ 1.17613\times 10^{-2}  $ & $ 0.123312 $ &   $ 0.394704 $  \\
    \textbf{He:Ne laser}&$10^{8}$ & $ 6.3495\times 10^{-3}  $ & $ 1.75833\times 10^{-4}  $  & $ 1.26278\times 10^{-2}  $ &  $ 5.83538\times 10^{-2}  $  \\
    $\omega=2\,(eV) $&$10^{9}$ & $ 6.4924\times 10^{-4}  $ & $ 1.85626\times 10^{-6}  $ & $ 1.2912\times 10^{-3}  $ & $ 6.2441\times 10^{-3}  $  \\
    &$10^{10}$ & $ 6.10791\times 10^{-5}  $ & $ 1.69926\times 10^{-8} $ & $ 1.21473\times 10^{-4}  $ &    $ 5.90086\times 10^{-4} $  \\
    &$10^{11}$ & $ 6.36515\times 10^{-6}  $ & $ 1.75454\times 10^{-10} $ & $ 1.26589\times 10^{-5}  $ &    $ 6.15229\times 10^{-5} $  \\
    &$10^{12}$ & $ 6.45132\times 10^{-7}  $ & $ 1.82467\times 10^{-12} $  & $ 1.28303\times 10^{-6}  $ &  $ 6.23585\times 10^{-6} $  \\
    &$10^{13}$ & $ 6.25258\times 10^{-8}  $ & $ 1.75634\times 10^{-14} $  & $ 1.2435\times 10^{-7}  $ &  $ 6.04378\times 10^{-7} $  \\
     \hline
\end{tabular}
\end{table*}
\normalsize
Table \ref{tab3} illustrates the angular distribution of the produced Higgs-strahlung as a function of the laser field strength for both $e^{+}e^{-}\rightarrow\mu^{+}\mu^{-}H$ and $e^{+}e^{-}\rightarrow\nu\bar{\nu}H$ and for two different known laser sources; i.e., the \textbf{Nd:YAG laser} and \textbf{He:Ne laser}. The number of exchanged photons between the electromagnetic field and both the initial ($n$) and final ($N$) particles is chosen as ranging from $-10$ to $+10$. Results presented in columns 3 and 5 correspond to the case where only the initial $e^{+}e^{-}$ beam is embedded in the laser field while columns 4 and 6 correspond to the case where both the initial and final particles are embedded in the laser field. For low laser field intensities, the laser sources doesn't influence the angular distribution of the produced Higgs boson regardless of the laser source type. However, whenever the laser field strength overcomes a certain threshold value, the angular distribution begins to decreases progressively as much as this strength increases. This laser field strength threshold value is equal to $\varepsilon=10^{5}\, V.cm^{-1}$ for both processes in the case where $\omega=1.17\,eV$. Whereas, for $\omega=2\,eV$, this threshold value is $\varepsilon=10^{6}\, V.cm^{-1}$. Consequently, the effect of the laser field appears at low laser strength for high laser sources frequency. 
This effect is interpreted by the fact that the produced pair of muons acquire an effective mass inside an external field with high strengths and low frequencies.
We compare results obtained in the case where only the initial beam is embedded in a laser field with those where both initial and final particles are considered in an electromagnetic field. We remark that the laser-assisted cross-section decreases in the presence of a laser field for both processes. 
However, for the process $e^{+}e^{-}\rightarrow\mu^{+}\mu^{-}H$ and if all initial and final particles are inside the electromagnetic field, the differential cross-section decreases strongly as compared to that in the case where only the incoming $e^{+}e^{-}$ beam is dressed by a laser field. Moreover, the laser effect is as strong as the laser field strength increases. On contrary, for the process, $e^{+}e^{-}\rightarrow\nu\bar{\nu}H$, the differential cross-section is enhanced in the case where all the particles that contribute to the process are embedded in the electromagnetic field.
The reader may ask why the angular distribution of the Higgs boson changes by embedding the produced Higgs boson and the produced pair of neutrinos $\nu\bar{\nu}H$ in the laser field though they are neutral. 
The response is that the $Z$-boson branching ratio in equation (\ref{7}) depends also on $\Gamma^{N}_{tot}$ which is affected inside the laser field as it contains the term $\Gamma_{N_{({Z\rightarrow \mu^{+}\mu^{-}})}}$.
In addition, it is proven in \cite{dec3} that due to the increase in the effective mass that fermions acquire inside the electromagnetic field, the $Z$-boson could only decay invisibly into neutrinos, and its probability to decay into any other pair of charged fermions becomes negligible.
It is well known that laser sources differ by their frequencies. Therefore, we have plotted in figure $\ref{fig4}$, the Higgs boson angular distribution for different laser field frequencies.
\begin{figure*}[t!]
  \centering
      \includegraphics[scale=0.5]{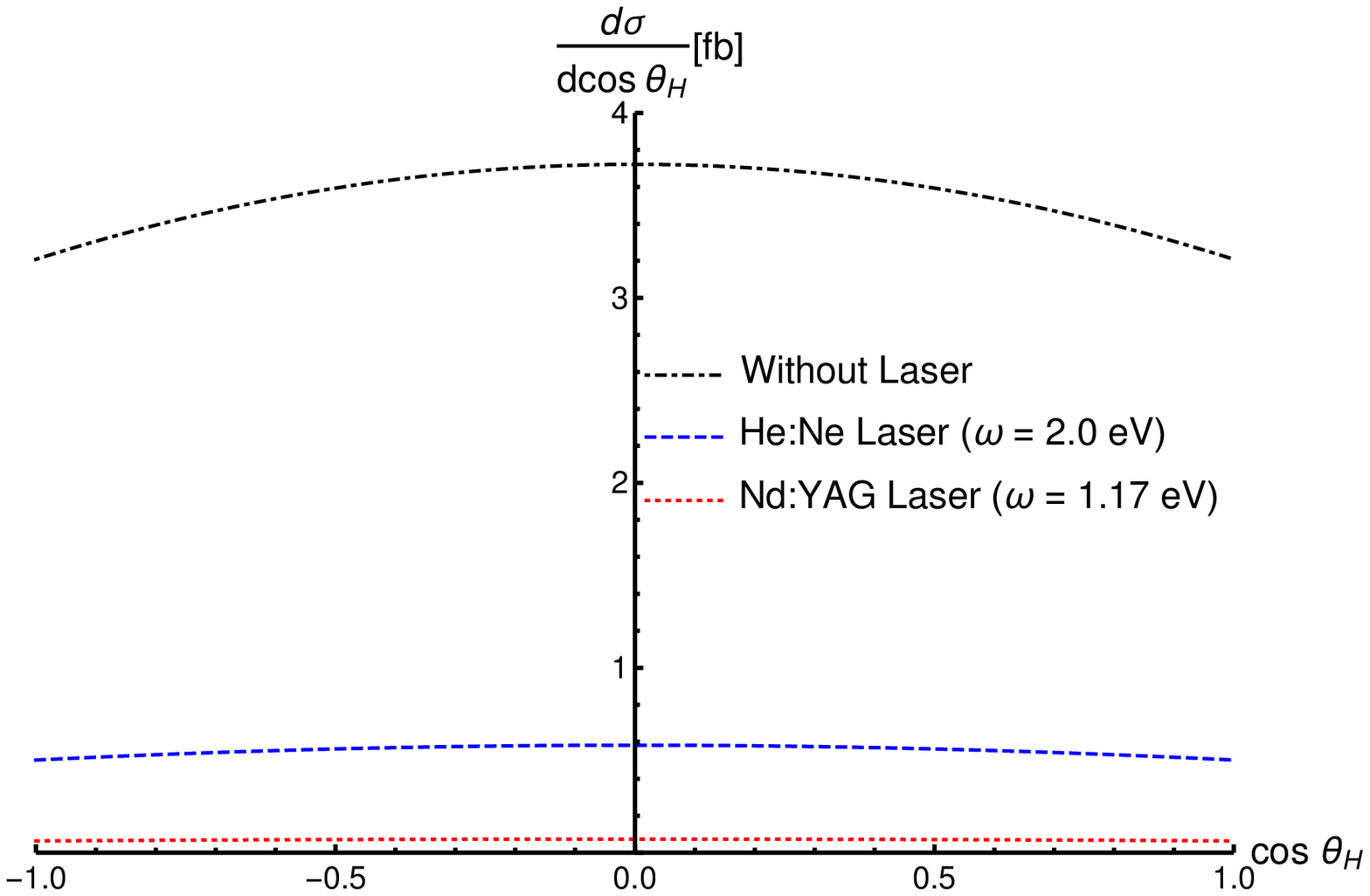}\hfill
      \includegraphics[scale=0.5]{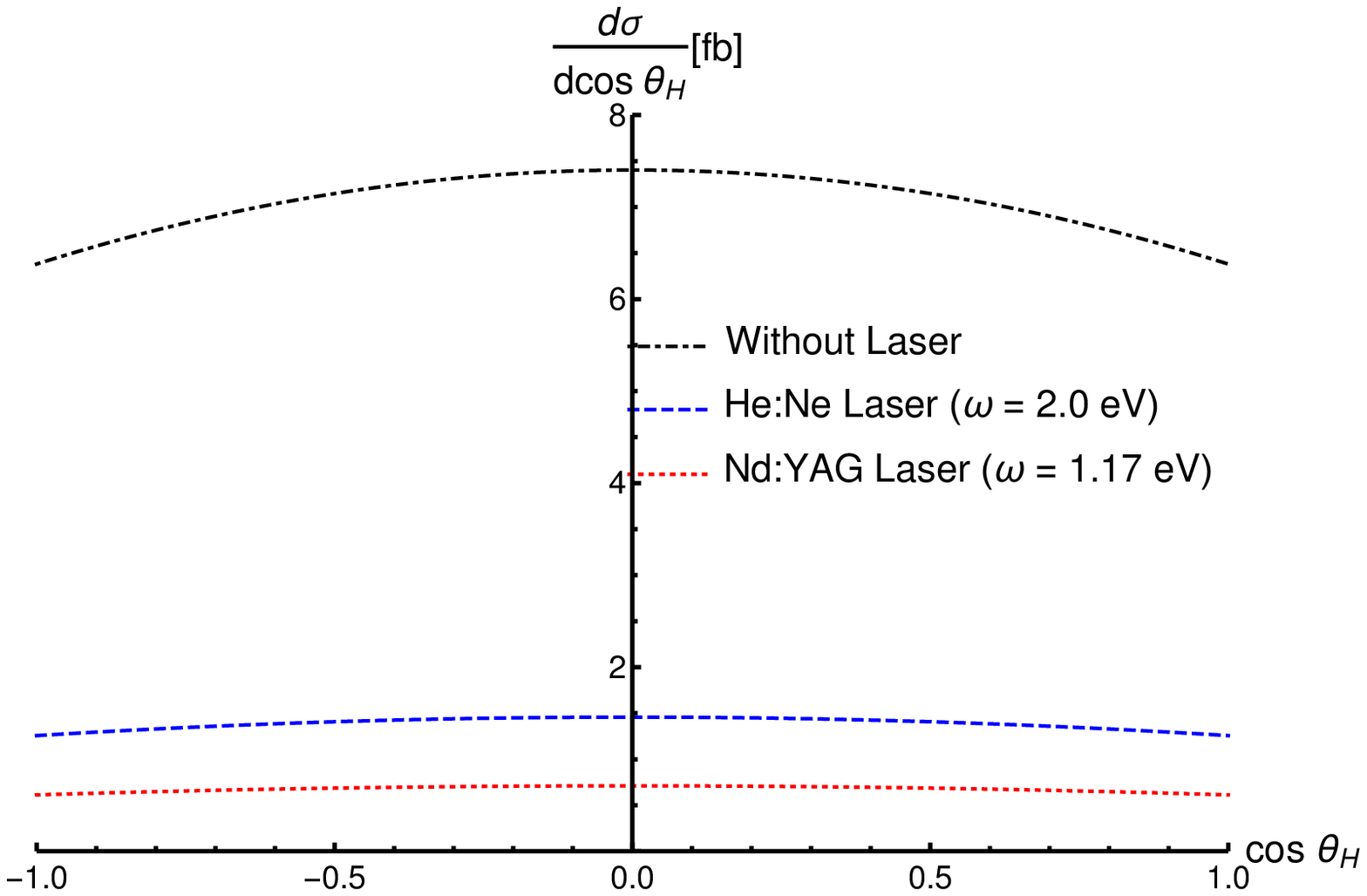}\par\medskip
        \caption{Angular distribution of the Higgs boson production inside the laser field at $\sqrt{s}=240\,GeV$ for different frequencies and for both $e^{+}e^{-}\rightarrow\mu^{+}\mu^{-}H$ (left) and $e^{+}e^{-}\rightarrow\nu\bar{\nu}H$ (right). We have chosen: $\varepsilon=10^{6}\, V.cm^{-1}$, $\sqrt{s}=240\,GeV$ and $n=N=\pm10$.}
        \label{fig4}
\end{figure*}
According to figure $\ref{fig4}$, the differential cross-section of both processes decreases inside the laser field as far as we decrease its frequency. We remark that this figure is in agreement with the previous ones as the differential cross-section decreases inside the electromagnetic field of circular polarization. However, the small order of magnitude of the laser-assisted differential cross-section is due to the fact that we have chosen the numbers of exchanged photons $n$ and $N$ as ranging from $-10$ to $+10$. These numbers of exchanged photons are very small as compared to the corresponding cutoffs, and the choice is made to avoid extensive numerical calculation. An important point to mention here is that the frequency of the laser source has a great impact on the cross-section. In addition, for both processes, the differential cross-section decreases as far as we decrease the laser frequency.
\section{Conclusion}\label{conclusion}
In this paper, we have considered the process of Higgs strahlung production in association with a pair of muons or neutrinos at the lowest order in the presence of a circularly polarized laser field. 
The theoretical calculation of the differential cross-section is done within the framework of the scattering-matrix approach by describing the fermions inside the laser field by Dirac-Volkov states and by using the narrow width approximation. 
Then, the analytical laser-assisted differential cross-section is analyzed numerically for both processes in two cases; in the first case, we have only dressed the initial particles while both initial and final particles are considered inside the laser field in the second case. As a result, the effect of the laser on the differential cross-section of $e^{+}e^{-}\rightarrow\mu^{+}\mu^{-}H$ is stronger in the second case. However, for the process $e^{+}e^{-}\rightarrow\nu\bar{\nu}H$, its differential cross-section is enhanced by embedding both initial and final particles in the electromagnetic field.

\end{document}